\pdfoutput=1
\documentclass{article}
\usepackage[margin=1in]{geometry}

\usepackage[final]{microtype}
\usepackage{subcaption}
\usepackage[x11colors]{xcolor}
\usepackage{tikz}
\usetikzlibrary{shapes,decorations.pathreplacing,decorations.pathmorphing,arrows,positioning,matrix,calc}
\tikzset{>=stealth}
\usepackage{amsmath,amssymb,amsthm,mathtools}
\usepackage[scr=boondox]{mathalfa}
\usepackage[colorlinks]{hyperref}

\usepackage{listings}

\tikzset{
	block/.style = {draw, rectangle, 
		minimum height=1cm, 
		minimum width=2cm},
	input/.style = {coordinate,node distance=1cm},
	output/.style = {coordinate,node distance=4cm},
	arrow/.style={draw, -latex,node distance=2cm},
	pinstyle/.style = {pin edge={latex-, black,node distance=2cm}},
	sum/.style = {draw, circle, node distance=1cm}
}

\numberwithin{equation}{section}

\theoremstyle{definition}

\theoremstyle{remark}

\begin{document}
\title{Correlations of Multi-input Monero Transactions}
\author{Nathan Borggren and Lihan Yao}
\date{Geometric Data Analytics, Inc., Durham, NC 27701}
\maketitle

\begin{abstract}
A variety of correlations are detected in the Monero blockchain.  
The joint distribution of the time-since-last-transaction between elements of pairs of RingCTs is enhanced
in comparison with the product of the marginal distributions.  
Similarly there is an enhancement in the joint distribution of the hour timestamps between the same pairs.
Lastly, we find another enhancement when the correlation is measured between the hour timestamps of the transaction itself and the elements of the RingCTs.
We calculate some adjustments to the probabilities of which input in a RingCT is real, providing an additional heuristic to 
denoising the Monero blockchain. 

\end{abstract}

\section{Introduction}

Privacy issues with respect to the pseudo-anonymity of Bitcoin-like blockchains were understood since its origin \cite{Nakamoto}.
Techniques were developed to exploit these vulnerabilities by interacting directly with entities and using the gained information to tag these entities \cite{Meiklejohn2013}.
Indeed, at this stage of Bitcoin's development, large scale datasets are freely available e.g. \cite{walletexplorer} and have been used
for machine learning, trading, and economics studies \cite{deep_borggren_2017}.
Topological methods that take advantage of the graph-nature of blockchains have also been combined with machine learning to characterize exchanges \cite{Ranshous2017, ss_borggren_2018} and identify malware \cite{DBLP:journals/corr/abs-1906-07852}.
Any attempt to capture all such deanonymization efforts will be partial and unsatisfactory but for a deeper review consider \cite{Conti}.

As a result of these privacy limitations new coins have emerged and gained in popularity.  
Monero, ZCash, and Grin have added to the bitcoin protocol or have developed entirely new cryptographic techniques to address these issues.
Analysis of the activity in 2018 of the cryptocurrency exchange ShapeShift showed a retreat from Bitcoin towards privacy-centric coins \cite{ss_borggren_2018}.
The Shapeshift terms of service itself acknowledges the irony of using this exchange to move into these coins for privacy purposes and advises against it.
Monero developers have also warned users of dangers in overconfidence in the privacy offered and continuously seek improvements to their methods.

Nonetheless, it is apparent that the ShapeShift exchange and Monero in particular have been used to attempt laundering of illicit activities.  
Regardless of the activity being illicit or not, through ShapeShift or not, heuristics and techniques have been developed to analyze Monero exposing users
to exposure of their privacy and providing law enforcement analytic tools for deeper investigation.
This present work will introduce and develop an additional heuristic that can be used to aide in tracing real transactions through the Monero blockchain.

\section{Correlations}

When a Monero transaction aggregates coins from multiple inputs the RingCT feature of Monero
mixes the real transaction with currently 10 past transactions randomly chosen from the blockchain.
Monero uses a draw from a known gamma distribution with parameters given by $shape = 19.28$ and $scale = \frac{1.}{1.61}$
 to repeatedly pick times and
then a transaction near that time is found in the blockchain and chosen for mixing.  (see lines 132-133 at \cite{xmr_github})

The distribution of inputs then can be expressed as a mixture $P(t) = P(t|\theta_{real})+P(t|\theta_{fake})$.
However we hypothesize that any correlation that occurs can only come from pairs of real transactions inputs, 
as draws from the fake distribution do not depend on the real input or previous draws from the fake distribution.
Thus correlations may be present that are revealing to which input is the real input. 

We seek then to measure empirically the correlation: 

\begin{equation}
C(t_1, t_2) = \frac{P(t_1, t_2)}{P(t_1)P(t_2)}.
\end{equation}

For example, consider a transaction with two RingCT inputs.  There are then 121 pairs of inputs for which we fill our 2d histogram with the measured $t_i$ times giving us the empirical joint distribution.
We can also collect the times of the 22 involved inputs to get the marginal distributions.
For transactions with more than two inputs, this procedure can be repeated for each pair of inputs.

\begin{figure}\
	\centering
	\begin{subfigure}[h]{0.49\linewidth}
	\includegraphics[width=\linewidth]{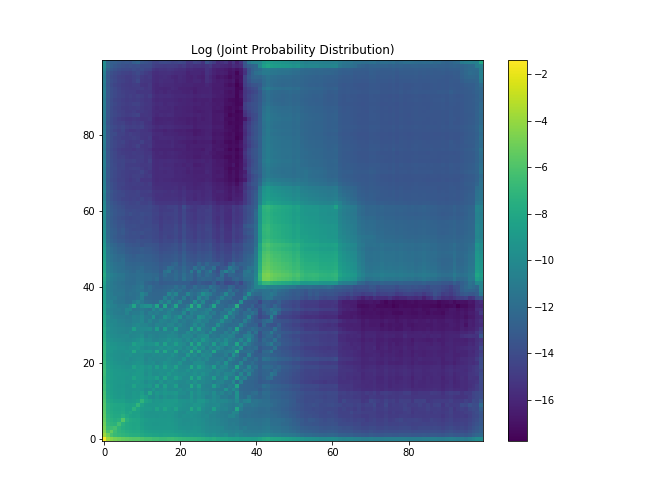}
	\caption{Log of the Joint Distribution, $P(t_1, t_2)$ }
	\end{subfigure}
	\hfill
	\begin{subfigure}[h]{0.49\linewidth}
	\includegraphics[width=\linewidth]{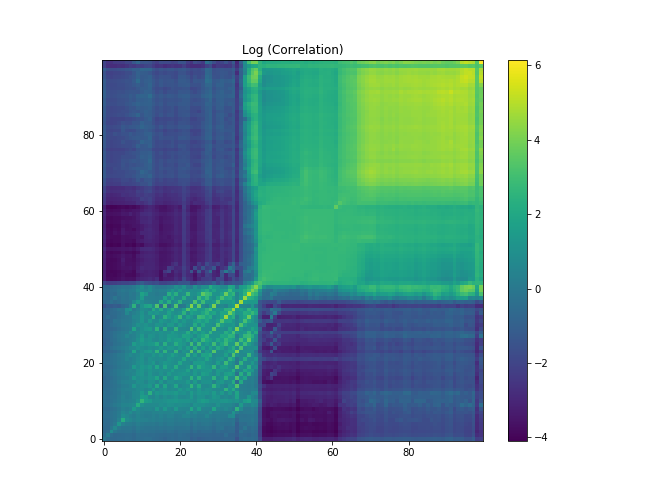}
	\caption{Log of the Correlation, $C(t_1, t_2)$}
	\end{subfigure}%
	\caption{Empirical Joint and Correlation functions, a bin is approximately 16 days wide.}
	\label{fig:foreground}
\end{figure}

Correlations are measured this way over 150000 blocks, 1650000 to 1800000, and shown in fig. \ref{fig:foreground}.



\subsection{Mixed-event correlations}

We would like to insure that any correlation arising is a result of the signer of the transaction
having authority over the pair of inputs.  However, this is not guaranteed.  
To mod out the background correlations that can emerge from other sources (low-sampling) we
build an empirical distribution $C_{background}(t_1, t_2)$ where $t_1$ and $t_2$ are now from the inputs of \textit{different}
transactions.  The mixture component, $P(t|\theta_{real})$, is itself really a mixture of all the individual users. 
We would like for the correlation to come from a single individual user, not the whole collective.  By constructing
this background correlation we can reveal the correlation information that occurs between two random participants of the network.

\begin{figure}\
	\centering
	\begin{subfigure}[h]{0.49\linewidth}
	\includegraphics[width=\linewidth]{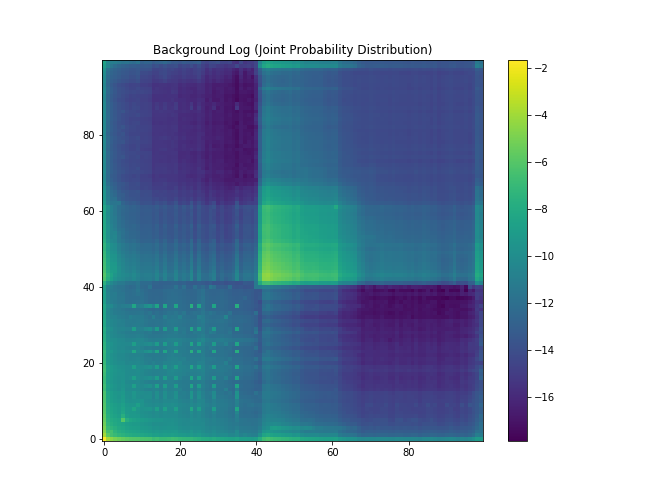}
	\caption{Log of the background Joint Distribution, $P(t_1, t_2)$. }
	\end{subfigure}
	\hfill
	\begin{subfigure}[h]{0.49\linewidth}
	\includegraphics[width=\linewidth]{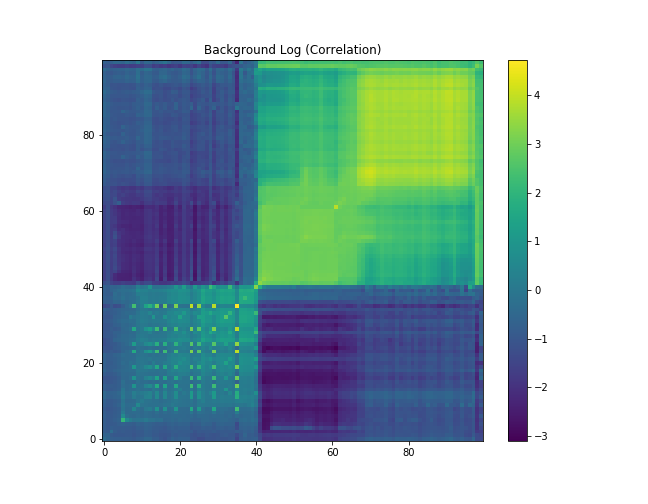}
	\caption{Log of the background Correlation, $C(t_1, t_2)$.}
	\end{subfigure}%
	\caption{Background Joint and Correlation functions.  Again a bin is approximately 16 days wide.
	While correlations are maintained from transaction volume, the same-wallet correlations have been removed by correlating across transactions.}
	\label{fig:background}
\end{figure}

By close inspection of figs \ref{fig:foreground} and \ref{fig:background} one can see that 
there is indeed structure in the foreground not present in the background, seen for example along the diagonal from the lower left to top right
wherein the foreground the high correlations remain high over larger time scales.

The expression for correlation that enhances these differences is $\frac{C_{foreground}(t_1,t_2)}{C_{background}(t_1,t_2)}$ and shown in fig \ref{fig:corrmod}.  
The fact that this expression is non-zero implies different individuals have different transactional habits, 
providing an opportunity to look for signal associated to particular individuals habits.  
Such an analysis has been carried out for bitcoin \cite{Monaco2015}, but has not been done for Monero.

\begin{figure}[h]
	\centering
	\includegraphics[width=.8\linewidth]{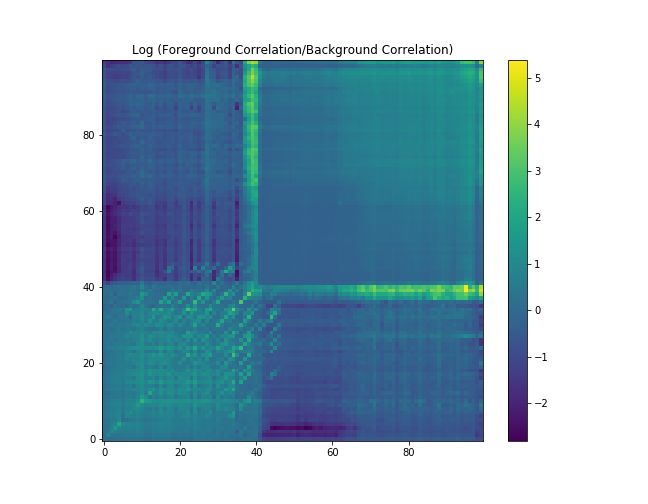}
	\caption{Log of $\frac{C_{foreground}(t_1,t_2)}{C_{background}(t_1,t_2)}$.  A bin is approximately 16 days wide.}	
	\label{fig:corrmod}
\end{figure}

\subsection{CDF-binned Histograms}

The bins used for the first stage of this analysis were rather wide; one bin is 16 days.  
However there is interesting structure emerging already.  Bin `40' is about October of 2017, the beginning of one of Bitcoin's remarkable bull runs.
The high correlations for that time period are likely an artifact of the simple fact that an anomalous
number of transactions come from that time period simply because of the volume of transactions occuring during that bull run.
We are using the assumption that, generally, the price of Monero tracks the price of Bitcoin,
which we use as a proxy for the whole cryptocurrency market.

We remind the reader that these are log plots, so that these correlations are incredibly large.  

The gamma distribution used for mixins as well as the apparent-real-use transaction behavior in Monero are very heavily weighted towards short time scales.
This means we are histogramming over a lot of potentially revealing timescales.  
One bin in our histogram corresponds to 16 days, which is around 11000 monero blocks, and a typical transaction will have many mixins and real contributions 
from this windowing.  

Thus, despite this remarkable correlation structure, at this stage the majority of pairs are filling
the same bin, the lowest left at the origin, so more effort needs to be made to use this fact to 
reveal what members of the rings are real and which are fake.

Transaction volume is also far from constant, so times of high volume require a large 
number of mixins as well, so it is possible there are high correlations between times of large transaction volume. 
Our background correlation in principle could account for that, but we had run the background for the same time range as the foreground.

Monero test networks, where we control the agents and their transaction behavior are being designed, that allow us to account for these sources or avoid them altogether.

\begin{figure}\
	\centering
	\begin{subfigure}[h]{0.49\linewidth}
	\includegraphics[width=\linewidth]{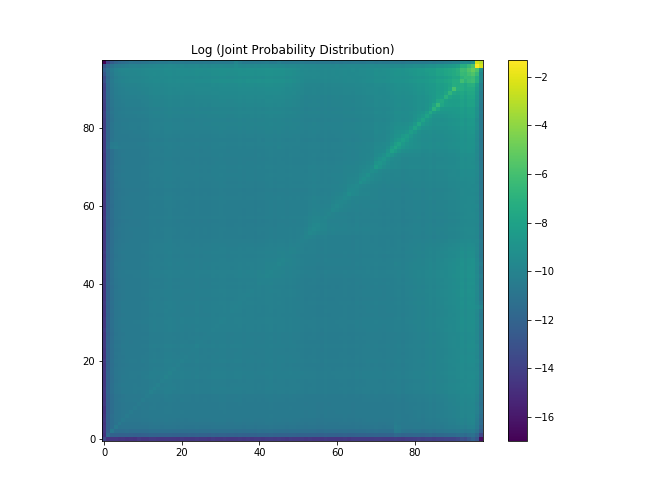}
	\caption{Log of the cdf-binned Joint Distribution, $P(t_1, t_2)$. }
	\end{subfigure}
	\hfill
	\begin{subfigure}[h]{0.49\linewidth}
	\includegraphics[width=\linewidth]{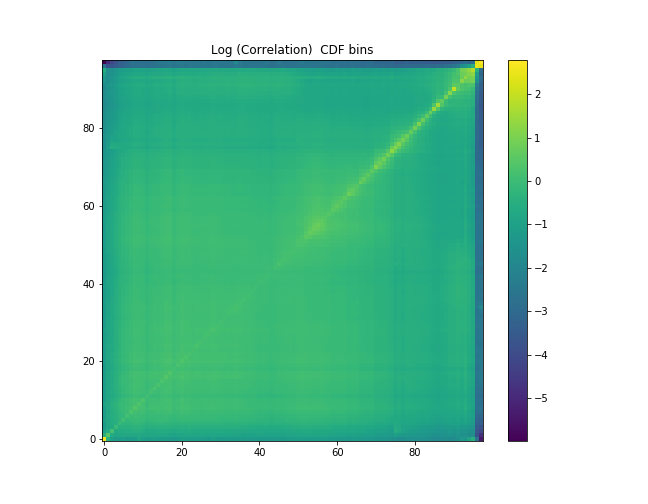}
	\caption{Log of the cdf-binned Correlation, $C(t_1, t_2)$.}
	\end{subfigure}%
	\caption{CDF binned Joint and Correlation functions. }
	\label{fig:cdf_background}
\end{figure}

We have repeated this procedure using bins with time edges computed from the $\Gamma$ cdf function so that 
each bin is equally likely to have a transaction in it.  Those results are shown in \ref{fig:cdf_background}. 

\begin{figure}[h]
	\centering
	\includegraphics[width=.8\linewidth]{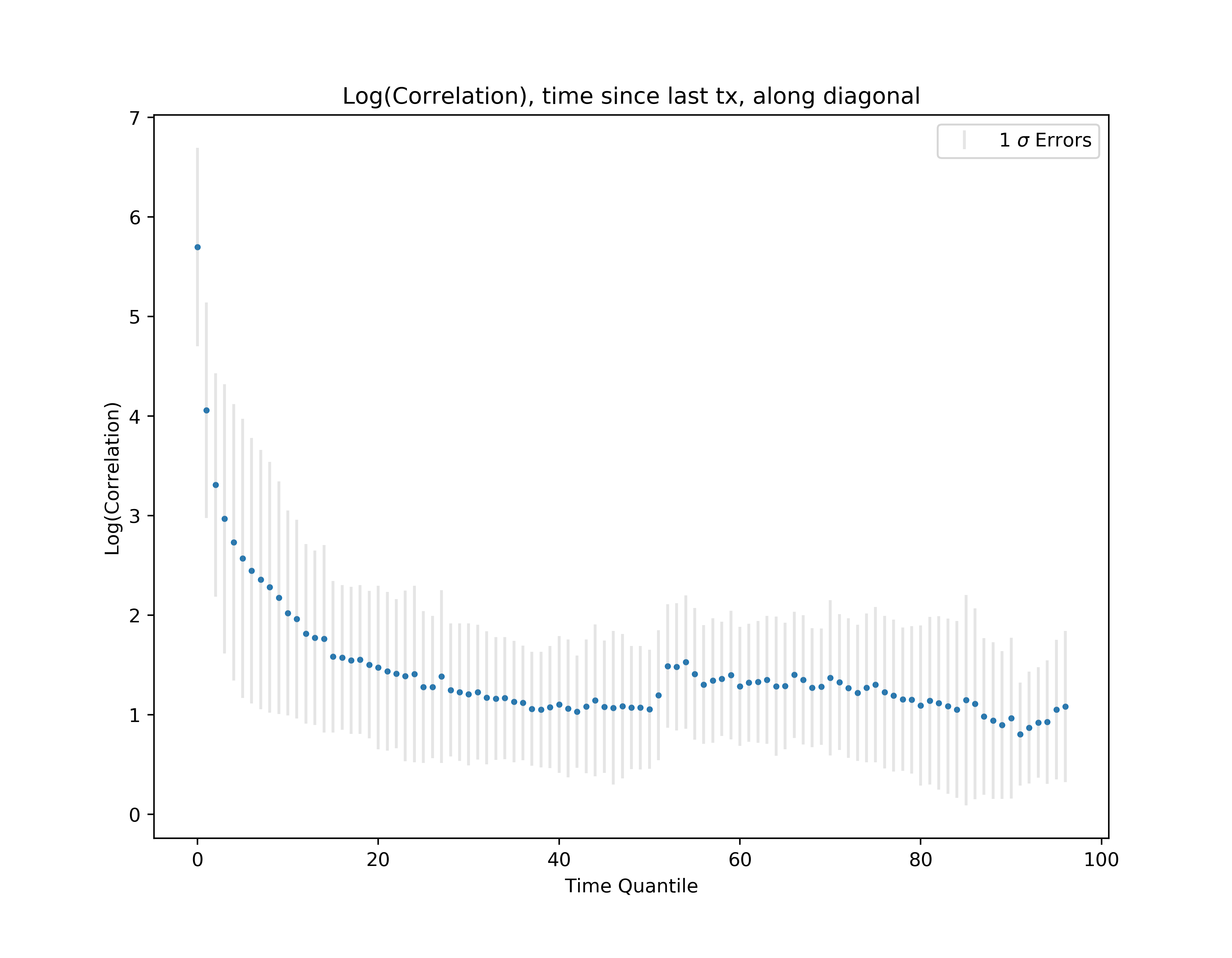}
	\caption{A 1-d figure showing the correlations along the diagonal, in which timestamps are contemporaneous.
	Error bars are computed by repeating the correlation calculation every 10000 blocks.}	\label{fig:cdfdiag}
\end{figure}

\subsection{Time-of-Day Correlations between rings}

\begin{figure}\
	\centering
	\begin{subfigure}[h]{0.49\linewidth}
	\includegraphics[width=.95\linewidth]{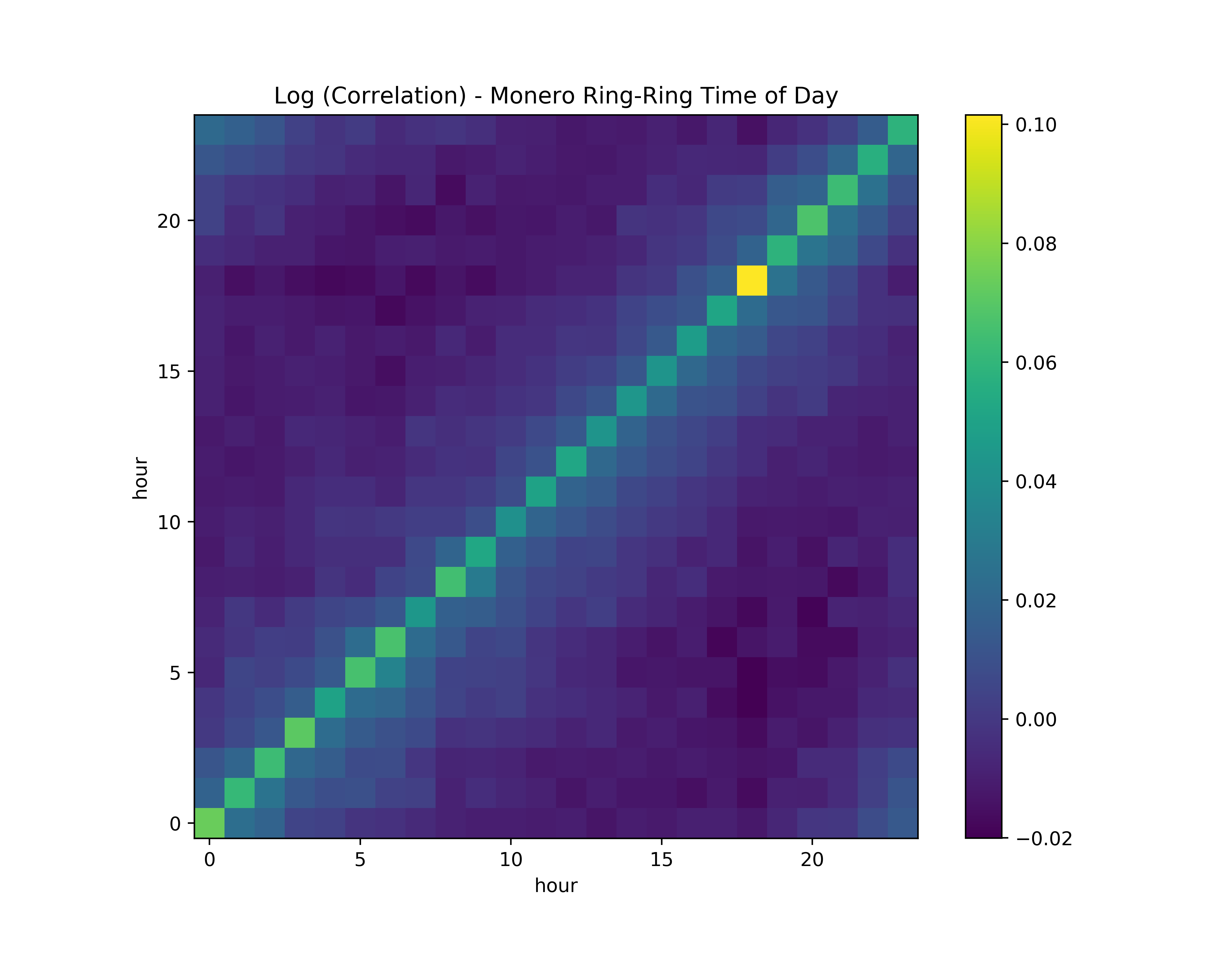}
	\caption{Log of $C_{ring-ring}(t_1,t_2)$.}	
	\label{fig:corrmod}
	\end{subfigure}
	\hfill
	\begin{subfigure}[h]{0.49\linewidth}
		\includegraphics[width=.95\linewidth]{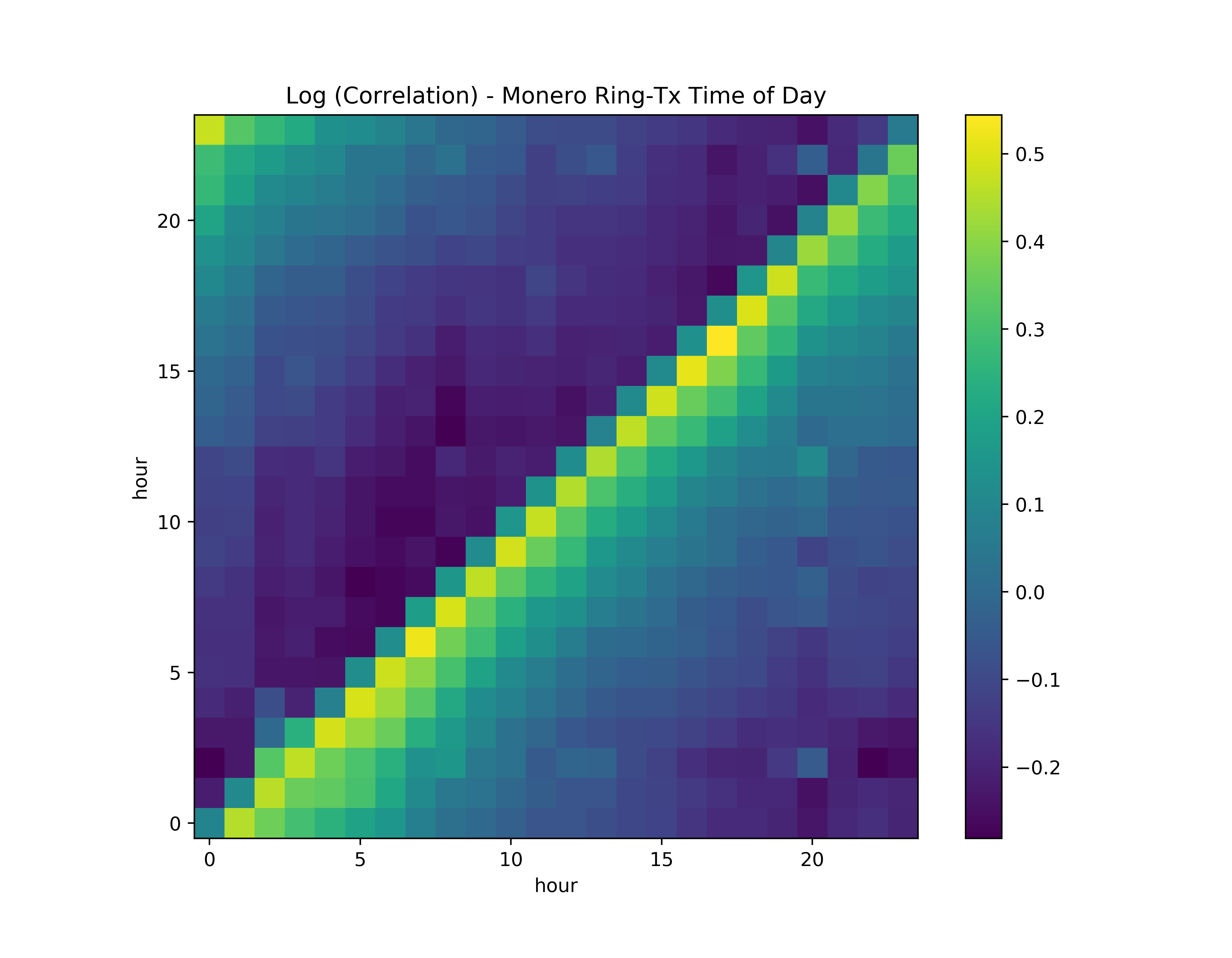}
		\caption{Log of $C_{ring-tx}(t_1,t_2)$.}		
	\end{subfigure}%
	\caption{Correlations arising from patterns-of-life emerge by histogramming the hour stamps. }
	\label{fig:ring_ring}
\label{fig:24hour}
\end{figure}

Next we investigate other candidate sources of correlations.  
For example, if a user transacts at the same times each day then a positive correlation would arise between the hour time stamps
of a ring and another ring from the presence of an abundance of transactions with particular hour time stamps.
Similarly, the correlation would arise as well for the same reason between the hour time stamps of a ring and the hour time stamp 
of the transaction itself.  Indeed this is what we measure and observe in fig. \ref{fig:24hour}.

\section{Translations into probabilities}

To understand the effect the correlations have on the probabilities of the ring-index being the real transaction let us first consider the Time-of-Day Correlations between
the hour of the transaction and the hour of the ring members.  

Let $C_{i-h, tx-h}$ denote the value of the correlation in the bin corresponding to the hour of input i and the tx-hour.
For example say $C_{i-h, tx-h}$ is 1.1 so that i-hour is 10 percent more likely to have occurred given tx-hour.
The probability increase must be offset by a decrease in probability in every other member of the ring.
All things being equal, we distribute the remaining probability over the $ringCT-1$ other possibilities.  
We construct the matrix equation to update the probabilities given the correlation.

\begin{equation}
\left[ \begin{array}{c} p_{1}' \\ p_{2}' \\ \dots \\ p_{ringCT}' \end{array} \right] = 
\begin{bmatrix} 
C_{1h, txh} & \frac{1}{ringCT-1}(1-C_{2h, txh}) & \dots & \frac{1}{ringCT-1}(1-C_{ringCTh, txh}) \\ 
\frac{1}{ringCT-1}(1-C_{1h, txh}) & C_{2h,txh} & \dots & \dots \\
\dots & \dots & \dots & \dots \\
\frac{1}{ringCT-1}(1-C_{1h, txh}) & \frac{1}{ringCT-1}(1-C_{2h, txh}) & \dots & C_{ringCTh,txh} 
\end{bmatrix} \times \left[ \begin{array}{c} p_{1} \\ p_{2} \\ \dots \\ p_{ringCT} \end{array} \right]
\end{equation}

\section{Conclusions}

We have shown that correlations in the Monero blockchain can be detected and can reveal information associated
to the real transaction within a ring.  
These correlations in future releases of Monero could be removed by adding a copula step to correlate
the spoofed transactions as well, but residue of this effect from past transactions would continue to propagate.

\newpage

\bibliographystyle{unsrt}
\bibliography{bitcoin}

\end{document}